\pdfoutput=1

\documentclass[11pt,twoside,a4paper,cmspaper,final,collab]{cms-tdr}
\def\svnVersion{185173}\def\cmsCernNo{2012-205}\def\cmsCernDate{\today}\def\cmsMessage{Submitted to Physical Review Letters}

\begin{document}\cmsNoteHeader{BPH-11-023}

\hyphenation{had-ron-i-za-tion}
\hyphenation{cal-or-i-me-ter}
\hyphenation{de-vices}

\newcommand{\UpsOne}{\PgUa}
\newcommand{\UpsTwo}{\PgUb}
\newcommand{\UpsThree}{\PgUc}
\newcommand{\Ups}{\ensuremath{\Upsilon}}
\newcommand{\UpsN}{\ensuremath{\Upsilon\mathrm{(nS)}}}
\newcommand{\costh}{\ensuremath{\cos\vartheta}}
\newcommand{\lth}{\ensuremath{\lambda_\vartheta}}
\newcommand{\lph}{\ensuremath{\lambda_\varphi}}
\newcommand{\ltp}{\ensuremath{\lambda_{\vartheta\varphi}}}
\newcommand{\ltilde}{\ensuremath{\tilde{\lambda}}}
\newcommand{\fbg}{\ensuremath{f_{\PB}}}
\def\Tiny{\fontsize{5pt}{5pt}\selectfont}

\newcommand{\mlab}[1]%
{\mbox{}\marginpar[\raggedleft\hspace{0pt}#1]{\raggedright\hspace{0pt}#1}}

\RCS$Revision: 162817 $
\RCS$HeadURL: svn+ssh://svn.cern.ch/reps/tdr2/papers/BPH-11-023/trunk/BPH-11-023.tex $
\RCS$Id: BPH-11-023.tex 162817 2012-12-29 21:44:03Z alverson $
\newlength\cmsFigWidth
\ifthenelse{\boolean{cms@external}}{\setlength\cmsFigWidth{0.85\columnwidth}}{\setlength\cmsFigWidth{0.4\textwidth}}
\ifthenelse{\boolean{cms@external}}{\providecommand{\cmsLeft}{top}}{\providecommand{\cmsLeft}{left}}
\ifthenelse{\boolean{cms@external}}{\providecommand{\cmsRight}{bottom}}{\providecommand{\cmsRight}{right}}
\cmsNoteHeader{BPH-11-023}
\title{\texorpdfstring{Measurement of the $\Upsilon$(1S), $\Upsilon$(2S), and
$\Upsilon$(3S) polarizations in $\Pp\Pp$ collisions at $\sqrt{s} = 7$\TeV}{Measurement of the Y1S, Y2S and Y3S polarizations in pp collisions at sqrt(s) = 7 TeV}}

\date{\today}

\abstract{The polarizations of the $\Upsilon$(1S), $\Upsilon$(2S), and $\Upsilon$(3S) mesons are
measured in proton-proton collisions at \mbox{$\sqrt{s} = 7$\TeV}, using a data sample
of $\Upsilon\mathrm{(nS)} \rightarrow \Pgmp\Pgmm$ decays
collected by the CMS experiment,
corresponding to an integrated luminosity of
4.9\fbinv. The dimuon decay angular distributions are analyzed in three different
polarization frames. The polarization parameters $\lambda_\vartheta$,
$\lambda_\varphi$, and $\lambda_{\vartheta\varphi}$, as well as the frame-invariant
quantity $\tilde{\lambda}$, are presented as a function of the $\Upsilon$(nS) transverse
momentum between 10 and 50\GeV, in the rapidity ranges $|y|<0.6$ and $0.6<|y|<1.2$.
No evidence of large transverse or longitudinal polarizations is seen in the
explored kinematic region.}

\hypersetup{%
pdfauthor={CMS Collaboration},%
pdftitle={Measurement of the Y1S, Y2S and Y3S polarizations in pp collisions at sqrt(s) = 7 TeV},%
pdfsubject={CMS},%
pdfkeywords={CMS, physics, quarkonium production, quarkonium polarization}}

\maketitle %maketitle comes after all the front information has been supplied

Studies of heavy-quarkonium production play a crucial role in the detailed
investigation of quantum chromodynamics (QCD), from the hard region, where an
expansion in the coupling constant is possible, to the soft region, dominated by
nonperturbative effects~\cite{bib:QWG-yellowreport}.
Given their high mass, heavy-quarkonium states are
approximately nonrelativistic systems, allowing the application of theoretical tools that
simplify and constrain the analyses of nonperturbative effects~\cite{bib:QWG}.
The differential cross sections of \JPsi\ and \Ups\ mesons produced at
Tevatron~\cite{Acosta:2004yw,Acosta:2001gv,Abazov:2005yc} and
LHC~\cite{Chatrchyan:2011kc,LHCb:2012aa, Khachatryan:2010zg} energies
can be reproduced by calculations based on nonrelativistic QCD (NRQCD)~\cite{bib:NRQCD},
dominated by ``color octet" production.
However, the corresponding predictions~\cite{NLO-NRQCD} of strong transverse polarizations (dominant angular
momentum component $J_z =\pm1$ with respect to the quarkonium momentum direction) are
in stark disagreement with the negligible polarizations measured for the \JPsi~\cite{PhysRevLett.99.132001}.
The \Ups\ satisfies the nonrelativistic approximation much better than the \JPsi,
making the \Ups\ polarization a more decisive test of NRQCD, especially at asymptotically large
transverse momentum, \pt. The existing measurements, however, are inconclusive, with the
CDF~\cite{PhysRevLett.108.151802} and D0~\cite{Abazov:2008aa} results in mutual
contradiction.

The polarization of the ($J^\mathrm{PC} = 1^{--}$) \Ups\ states can be measured through the study of the
angular distribution of the leptons produced in the $\Upsilon\to\Pgmp\Pgmm$
decay~\cite{bib:Faccioli-EPJC},
\begin{linenomath}
\begin{equation}
\label{eq:angular_distribution}
\ifthenelse{\boolean{cms@external}}{
\begin{split}
& W(\costh,\varphi|\vec{\lambda}) \, \propto \,
\frac{1}{(3 + \lambda_{\vartheta})}(1  +  \lambda_{\vartheta} \cos^2 \vartheta\, + \\
&
+ \lambda_{\varphi} \sin^2 \vartheta \cos 2 \varphi
+ \lambda_{\vartheta \varphi} \sin 2 \vartheta \cos \varphi ) \,,
\end{split}
}{%
W(\costh,\varphi|\vec{\lambda}) \, \propto \,
\frac{1}{(3 + \lambda_{\vartheta})}(1  +  \lambda_{\vartheta} \cos^2 \vartheta\,
+ \lambda_{\varphi} \sin^2 \vartheta \cos 2 \varphi
+ \lambda_{\vartheta \varphi} \sin 2 \vartheta \cos \varphi ) \,,
}
\end{equation}
\end{linenomath}
where $\vartheta$ and $\varphi$ are the polar and azimuthal angles, respectively, of the
$\Pgmp$ with respect to the $z$ axis of the chosen polarization frame.
As pointed out in
Refs.~\cite{bib:Faccioli-EPJC,PhysRevLett.102.151802,PhysRevLett.105.061601,
bib:Faccioli-PRD-FrameInv,PhysRevD.83.056008},
improved experimental measurements of quarkonium polarization require measuring all
the angular distribution parameters, $\vec{\lambda} = (\lth, \lph, \ltp)$, in different
polarization frames, as well as a frame-invariant polarization parameter,
$\tilde{\lambda} = (\lambda_\vartheta + 3 \, \lambda_\varphi) / (1-\lambda_\varphi)$.
This approach has already been followed in the \UpsN\ polarization analysis of
CDF~\cite{PhysRevLett.108.151802}, and in some recent theory
calculations~\cite{Baranov:2011ib}.

This Letter presents the first measurement of the polarizations of the \UpsN\
mesons produced in \Pp\Pp\ collisions at a center-of-mass energy of 7\TeV.
The analysis is based on a dimuon sample collected by the CMS experiment at the LHC,
corresponding to an integrated luminosity of
4.9\fbinv and containing %$\approx$\,
252\,000 \UpsOne, %$\approx$\,
94\,000 \UpsTwo, and %$\approx$\,
58\,000 \UpsThree\ mesons (after %applying
all selection criteria).

The analysis uses an unbinned likelihood approach, independent of assumptions on
the production kinematics.
The results are obtained in three
frames, with different directions of the quantization axis:
the center-of-mass helicity (HX) frame, where the polar axis coincides with the direction of the \Ups\
momentum; the Collins--Soper (CS) frame~\cite{bib:CS}, whose axis is the average of the two
beam directions in the \Ups\ rest frame; and the perpendicular helicity (PX) frame~\cite{Braaten:2008mz},
orthogonal to the CS frame.
The $y$ axis of the polarization frame is taken, in all cases, to be in the direction of the vector product
of the two beam directions, $\vec{P}_1 \times \vec{P}_2$ and $\vec{P}_2 \times \vec{P}_1$
for positive and negative rapidity, respectively.

The central feature of the CMS apparatus~\cite{Chatrchyan:2008zzk} is a superconducting
solenoid of 6\unit{m} internal diameter, providing a 3.8\unit{T} field.
The main subdetectors used in this analysis are the silicon tracker and the muon system.
The silicon tracker, composed of pixel and strip detector modules, is immersed in the magnetic
field and enables the measurement of charged-particle momenta over the pseudorapidity range
$|\eta| < 2.5$.
Muons are measured in the range $|\eta| < 2.4$ using gas-ionization detectors embedded
in the steel return yoke of the magnet and made using three technologies: drift tubes,
cathode strip chambers, and resistive plate chambers.
The events were collected using a two-level trigger system. %of CMS\@.
The first level consists of custom hardware processors and uses information from the muon
system to select events with two muons. The ``high-level trigger'' %processor farm
requires an
opposite-sign muon pair with invariant mass $8.5<M<11.5$\GeV, $|y| < 1.25$, $\pt > 5$ or 7\GeV
(depending on the instantaneous luminosity), and vertex fit $\chi^2$ probability greater than
0.5\%.

In the offline analysis, dimuons are formed by combining pairs of opposite-sign
muons (tracks in the silicon tracker matched to tracks in the muon detectors)
that satisfy several quality criteria, including the number of tracker hits, %in the silicon tracker,
the muon-track fit quality, and the vicinity of the track to the closest primary vertex
along the beam line.
The selected muons are required to
satisfy $|\eta| < 1.6$ and to have \pt above 4.5, 3.5, and
3.0\GeV for $|\eta| < 1.2$, $1.2 < |\eta| < 1.4$, and $1.4 < |\eta| < 1.6$, respectively, to ensure
accurately measured muon detection efficiencies.
Subsequent to the offline trigger %decision
confirmation, the combinatorial background
from uncorrelated muons is reduced by requiring a dimuon vertex fit
$\chi^2$ probability larger than 1.0\% and a distance between the dimuon vertex and the
closest primary vertex less than twice its uncertainty.
The analysis is performed in five dimuon \pt bins, of edges 10, 12, 16, 20, 30, and 50\GeV,
and two $|y|$ ranges, 0.0--0.6 and 0.6--1.2.

\begin{figure}[h]
\centering
\includegraphics[width=0.38\textwidth]{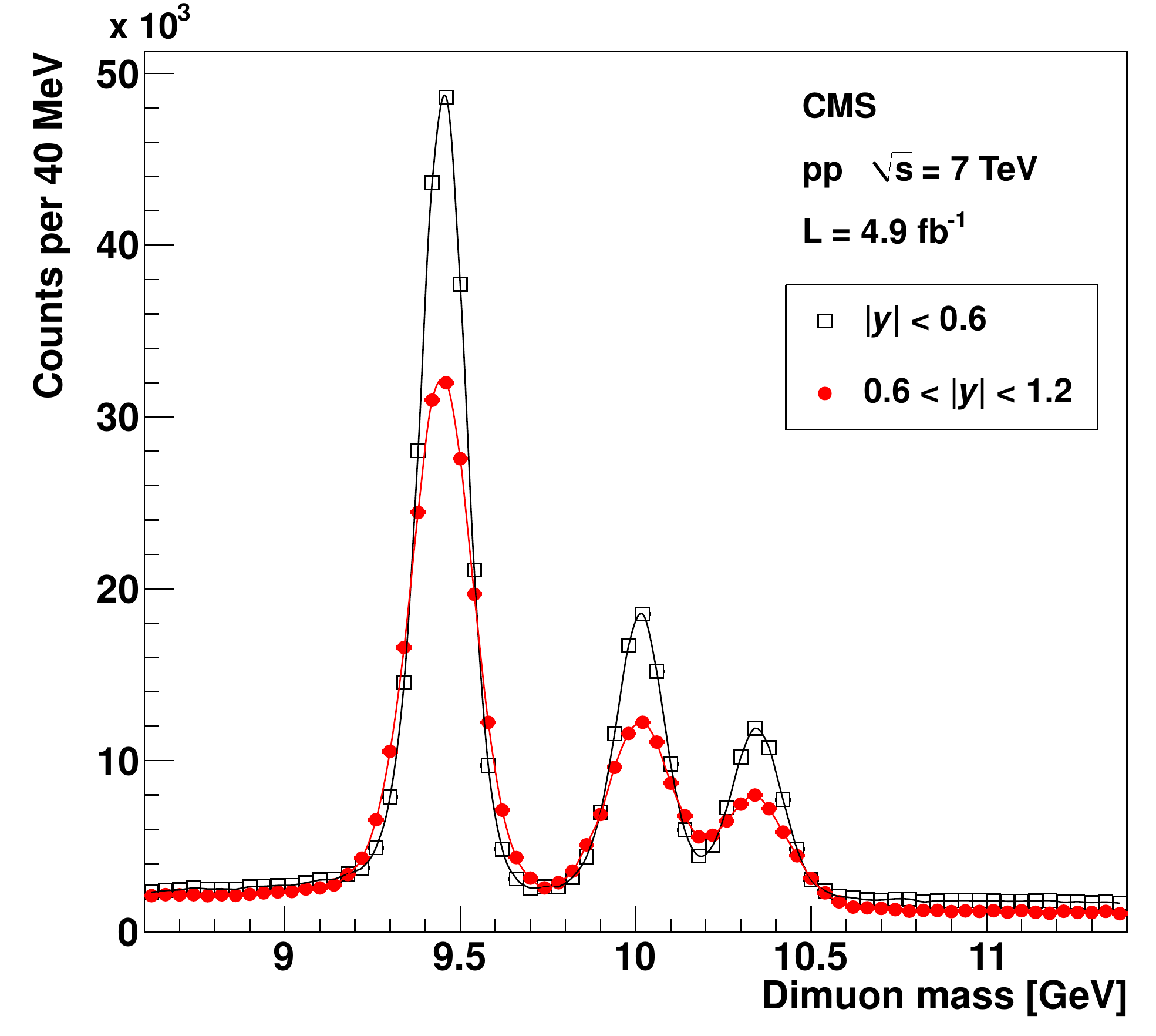}
\caption{Dimuon mass distributions in the \Ups\ region
for \mbox{$|y|<0.6$} (open squares) and
\mbox{$0.6<|y|<1.2$} (closed circles).}
\label{fig:dimuDist}
\end{figure}

The dimuon mass distribution, shown in Fig.~\ref{fig:dimuDist},
is well described by three Crystal-Ball functions~\cite{CrystalBall}
representing the \Ups\ peaks, and by a second-degree polynomial function
determined from the low- and high-mass sidebands,
located below the \UpsOne\ and above the \UpsThree, respectively.
The dimuon mass resolution is better than 70\MeV for $|y|<0.6$,
increasing to %$\approx$\,
95\MeV in the $0.6<|y|<1.2$ range, where the
\UpsTwo\ and \UpsThree\ peaks partially overlap.
Within a ${\pm}1$ standard deviation ($\sigma$) window around the
\UpsN\ masses, the cross-feed between the \UpsTwo\ and \UpsThree\ is below 4\%,
and the background fractions are 4--8\%, 9--18\%, and 12--28\% (increasing with
decreasing \pt), for the \UpsOne, \UpsTwo, and \UpsThree, respectively.

The single-muon detection efficiencies are measured with a ``tag-and-probe''
technique~\cite{Khachatryan:2010xn}
using event samples collected with dedicated triggers enriched in dimuons from \JPsi decays.
The trigger and reconstruction efficiencies must be accurately determined to avoid
biases on the angular distributions, which could mimic polarization effects.
The technique has been validated in the
fiducial region of the analysis with detailed Monte Carlo (MC) simulation studies.
The single-muon efficiencies are measured and parametrized as a function of \pt in
eight $\eta$ bins. Their uncertainties, reflecting the statistical precision of the calibration
samples and possible imperfections of the parametrization, contribute to the
systematic uncertainty on the final results.
The dimuon trigger and the selection criteria applied at the dimuon level could potentially
introduce muon-pair correlations, making the dimuon detection efficiencies different from
the product of the efficiencies of the two single muons. Detailed MC simulations show that
such correlations are essentially independent of $\costh$ and $\varphi$, in the phase space
selected for the measurement.
Residual effects are incorporated into the systematic uncertainty.

A fit to the dimuon mass distribution provides the fraction of background events, \fbg,
under each of the three \Ups\ mass peaks,
for a given definition of the signal region.
The angular distributions of these background events are modeled as weighted sums
of the distributions measured in the mass sidebands
(defined with negligible signal contamination),
with weights derived under the
assumption that they change monotonically with dimuon mass.
The background dimuons are subtracted on an event-by-event basis using the likelihood ratio $L_\mathrm{B} / L_\mathrm{(S+B)}$, where both likelihoods are functions of the variables $\pt, |y|, M, \costh$, and $\varphi$.
$L_\mathrm{B}$ is the likelihood of an event to be background, reflecting the background model, and $L_\mathrm{(S+B)}$ is its likelihood to be either signal or background, reflecting the distribution of the measured events.
A fraction \fbg\ of events distributed according to the $(\pt, |y|, M, \costh, \varphi)$ distribution of the background model is removed from the data sample.

The posterior probability distribution (PPD) for the average values of the \Ups\ polarization
parameters ($\vec{\lambda}$) inside a particular kinematic cell is then defined as a product
over the remaining (``signal-like'') events ($i$),
\begin{linenomath}
\begin{equation}
\label{eq:total_likelihood}
  \mathcal{P}(\vec{\lambda})= \prod_{i} \mathcal{E}(\vec{p}_1^{\,(i)},\vec{p}_2^{\,(i)})\,,
\end{equation}
\end{linenomath}
where $\mathcal{E}$ represents the event probability distribution as a function of the muon
momenta $\vec{p}_{1,2}$ in event $i$.
The priors are assumed to be uniform in the full parameter space.
Unlike most polarization analyses,
we do not use simulated $(\costh, \varphi)$ acceptance and efficiency maps,
averaged over all events in the considered kinematic cell.
Instead, the procedure exploits the efficiency measurement as a function of muon momenta,
attributing to each event a probability dependent on the full event kinematics
(not only $\costh$ and $\varphi$) and on the values of the polarization parameters.
The event probability is defined as
\begin{linenomath}
\begin{equation}
\label{eq:event_probability_definition}
\mathcal{E}(\vec{p}_1,\vec{p}_2)= \frac{1}{\mathcal{N}(\vec{\lambda})} \,
W(\costh,\varphi|\vec{\lambda}) \, \epsilon(\vec{p}_1,\vec{p}_2) \,,
\end{equation}
\end{linenomath}
where $\epsilon(\vec{p}_1,\vec{p}_2)$ is the detection efficiency.
The normalization factor $\mathcal{N}(\vec{\lambda})$ is calculated by integrating
$W \cdot \epsilon$ over $\costh$ and $\varphi$ uniformly,
using $(\pt, |y|, M)$ distributions determined from the background-subtracted data.

The background subtraction procedure is repeated 50 times to evaluate the statistical fluctuations associated with its random nature and the final PPD is obtained as the average of the 50 individual PPDs.

The analysis framework, including the effects of the detection efficiencies, has been
tested with pseudo-experiments based on simulated samples. Each test involves 50
pseudo-experiments and evaluates a specific systematic uncertainty. % of the measurement.
The pseudo-samples are individually generated and reconstructed, leading to statistically
independent determinations of the polarization parameters. The difference between the
median of the 50 results and the injected polarization parameters provides the systematic
uncertainty corresponding to the effect under study.
The reliability of the method to extract the signal polarization is evaluated for several
signal and background polarization scenarios. The influence of a possible residual bias from
muon or dimuon efficiencies, stemming from the tag-and-probe measurement precision or
from the efficiency parametrization, is evaluated by applying uncertainty-based %perturbations
changes to the efficiencies used in the extraction of the polarization parameters.
The monotonicity hypothesis in the
data-driven modeling of the background angular distribution under the \UpsN\
peaks has been tested by varying the signal region from $\pm 3$\,$\sigma$
to ${\pm}1\,\sigma$ around the \UpsN\ masses (with corresponding
corrections determined from a simple simulation of two-body decay kinematics). Despite significant changes in \fbg\
(from 40\% to 28\% for the \UpsThree\ at low \pt and $|y|<0.6$, for instance),
the results remain essentially identical for all three states.
Larger variations are observed by modifying the relative weights of the
low- and high-mass sidebands in the background model composition.
A conservative range of hypotheses is considered, such as assuming
that the background under the \UpsOne\ (\UpsThree) peak resembles exclusively the low-mass
(high-mass) sideband, or assuming that it is reproduced by an equal mixture of the two
sideband distributions.
While there is no dominant source of systematic uncertainty in the \UpsOne\ case, the total
systematic uncertainty of the \UpsTwo\ and \UpsThree\ states is dominated by the
background model uncertainty, especially at low \pt.
At high \pt, the statistical uncertainties dominate.
For example, the statistical uncertainties in \lth(PX) at $|y|<0.6$ for the
\UpsOne\ (\UpsThree) are of order 0.1 (0.2) at both low and high \pt;
the corresponding systematic uncertainties have a similar magnitude at low \pt
and are a factor of two (three) smaller at high \pt.

\begin{figure}[t]
\centering
\includegraphics[width= 0.38\textwidth]{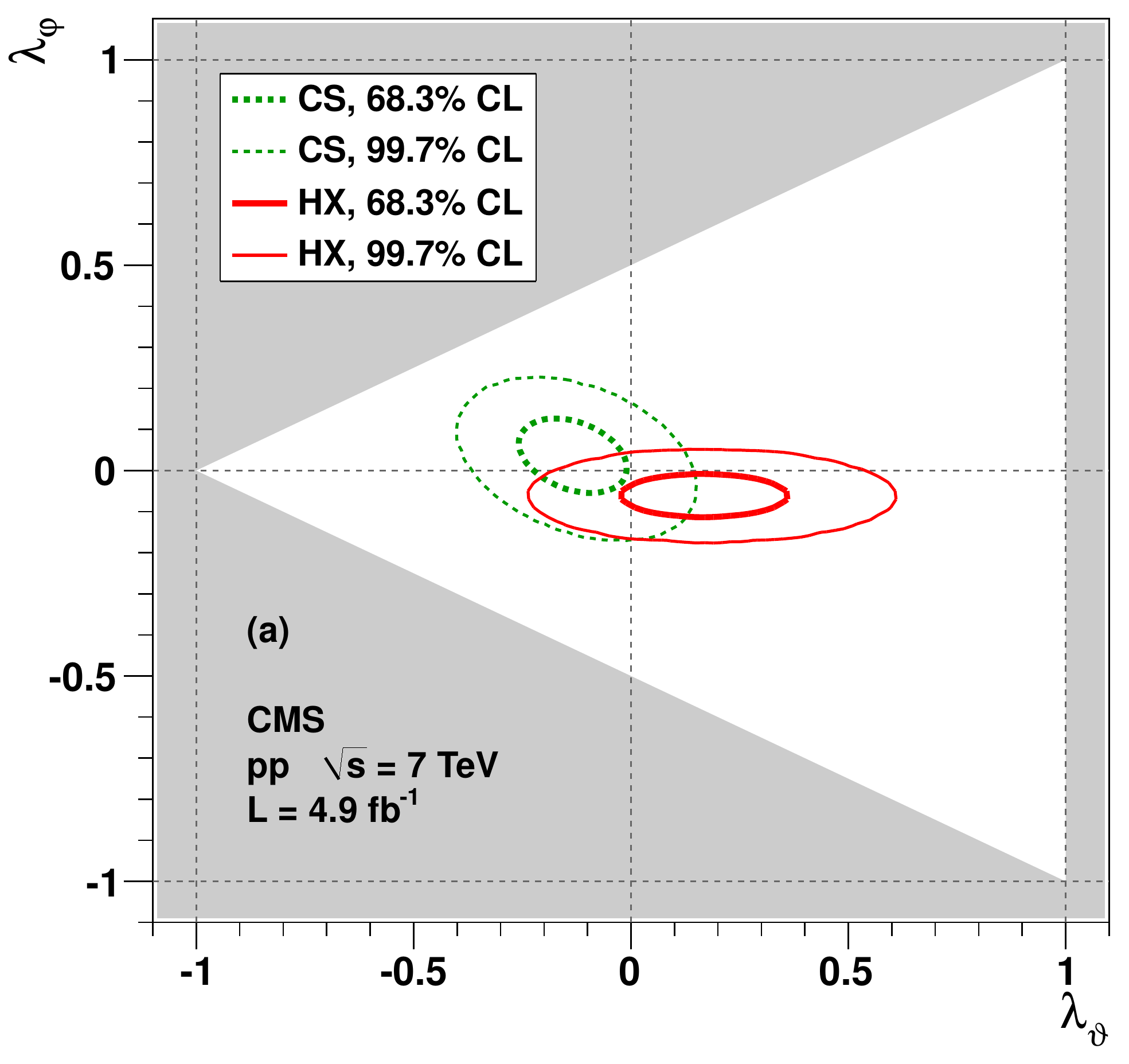}
\includegraphics[width= 0.38\textwidth]{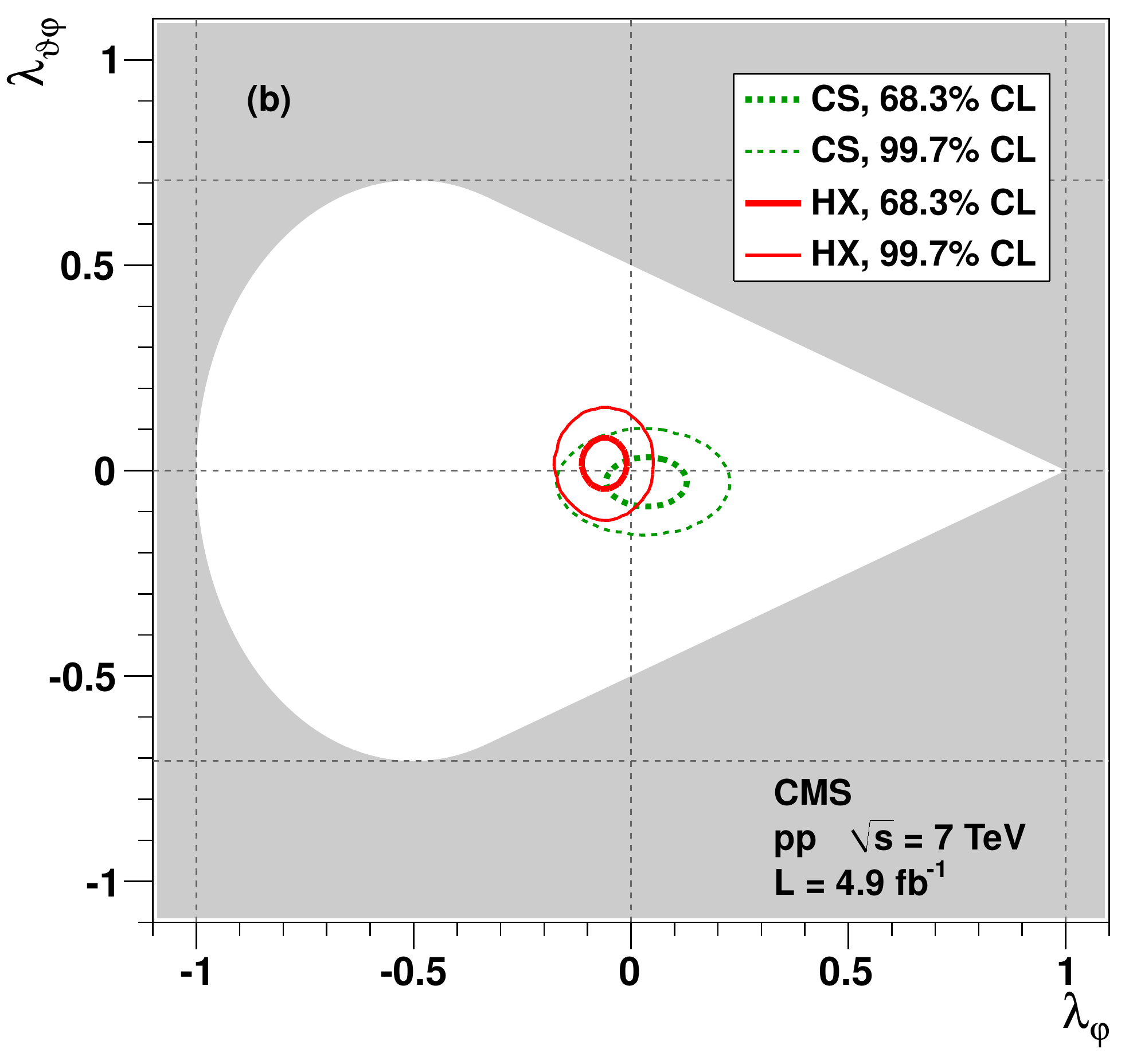}
\caption{Two-dimensional projections of the PPD in the \lph\ vs. \lth\ (a) and
\ltp\ vs. \lph\ (b) planes, for \UpsOne\ with \mbox{$|y|<0.6$} and
$30<\pt<50$\GeV. The 68.3\% and 99.7\% CL contours are shown for the CS and HX
frames. The shaded areas represent physically forbidden regions of parameter space~\cite{PhysRevD.83.056008}.}
\label{fig:2Dcorrelations}
\end{figure}

Each PPD is broadened by the effects of systematic uncertainties, which are included by convolution.
One- and two-dimensional projections of each final PPD are calculated by numerical
integration. %using the Metropolis--Hastings algorithm.
The highest posterior probability in
each one-dimensional
projection is used to estimate the best value of the associated polarization parameter.
Intervals
$[\lambda_1,\lambda_2]$
corresponding to a given
confidence level (CL), are calculated by identifying two regions of the parameter
space, $[-\infty,\lambda_1]$ and $[\lambda_2,\infty]$, each containing
$0.5\cdot(1-\mathrm{CL})$\% of the one-dimensional projection of the PPD\@.
Figure~\ref{fig:2Dcorrelations} shows two %-dimensional
projections of the final PPD for
the \UpsOne\ at $|y|<0.6$ and $30<\pt<50$\GeV,
displaying the 68.3\% and 99.7\% CL contours for the CS and HX frames.

\begin{figure*}[ht]
\centering
\includegraphics[width=0.77\linewidth]{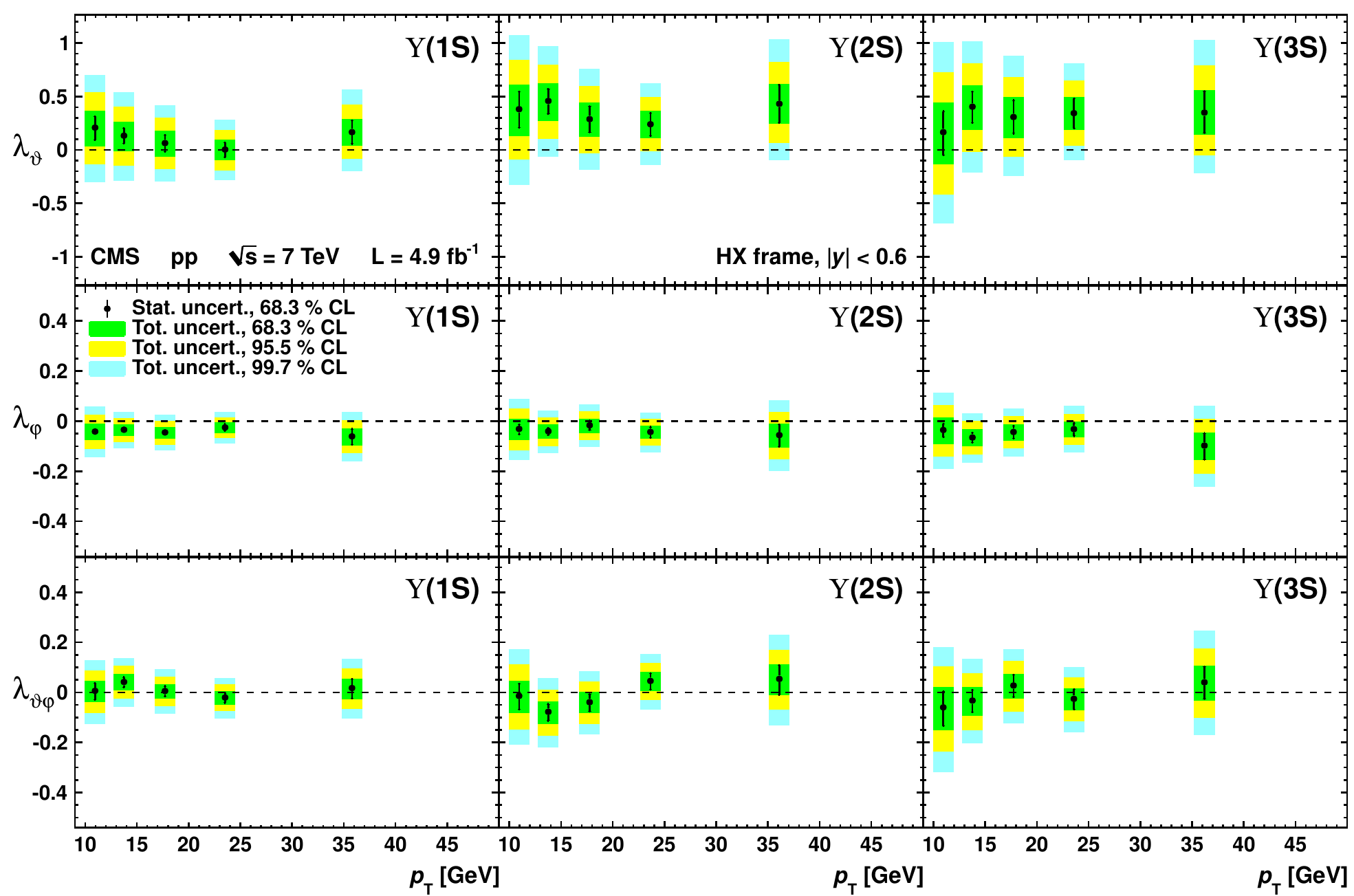}
\caption{Values of the \lth\ (top), \lph\ (middle), and \ltp\ (bottom) parameters for the \UpsOne\ (left), \UpsTwo\ (middle),
and \UpsThree\ (right), in the HX frame, as a function of the \Ups\ \pt for $|y|<0.6$. The
error bars indicate the 68.3\% CL interval when neglecting the systematic uncertainties.
The three bands represent the 68.3\%, 95.5\%, and 99.7\% CL intervals
of the total uncertainties. The points are placed at the average \pt of each
bin.}
\label{fig:mid-y}
\end{figure*}

\begin{figure*}[!t]
\centering
\includegraphics[width=0.77\linewidth]{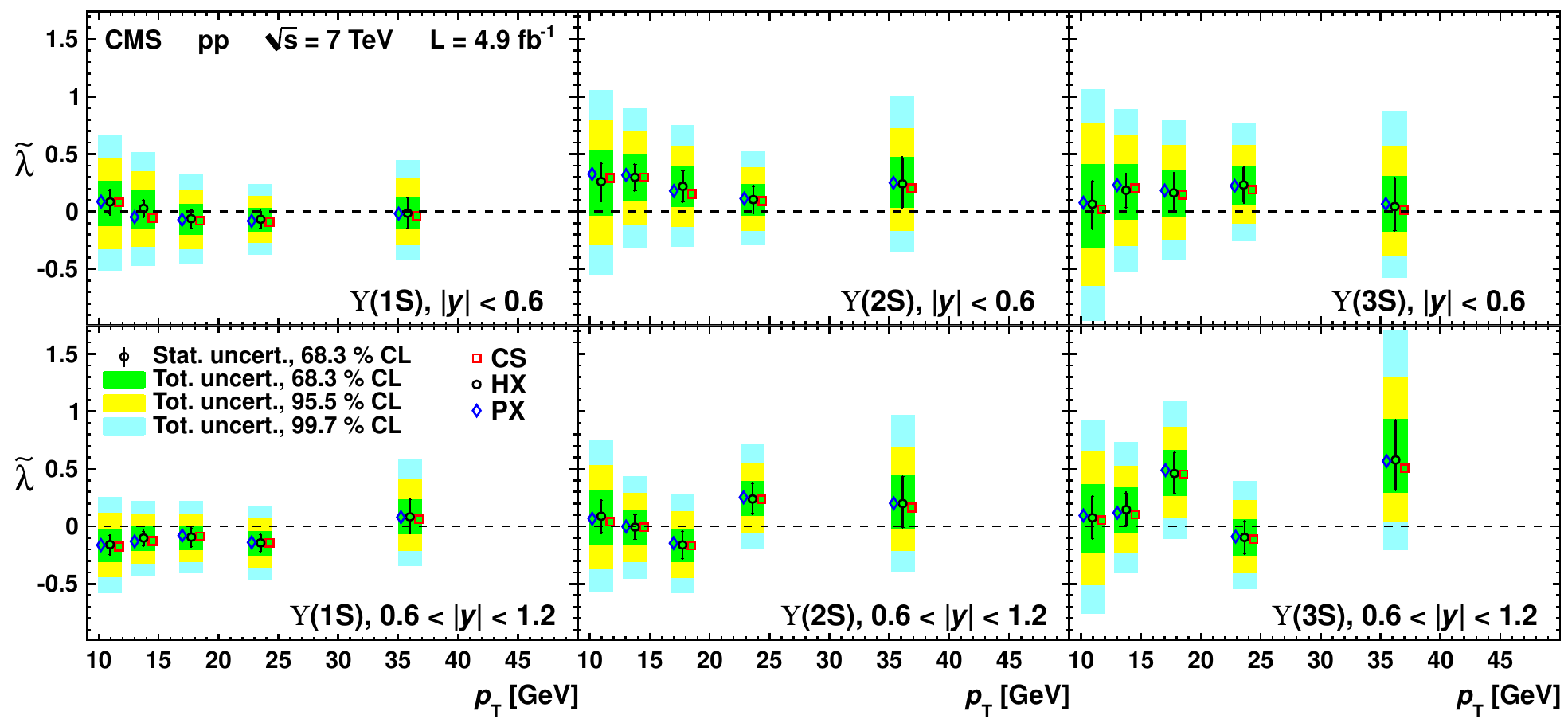}
\caption{Values of \ltilde\ for the \UpsOne, \UpsTwo, and \UpsThree\ states (left to right), in the
HX, CS, and PX frames, for the $|y|<0.6$ (\cmsLeft) and $0.6<|y|<1.2$ (\cmsRight) ranges. The
bands and error bars have the same meaning as in the previous figure.}
\label{fig:lambdatilde}
\end{figure*}

Figure~\ref{fig:mid-y} shows, for the rapidity range 0.0--0.6,
one-dimensional profiles (68.3\%, 95.5\%, and 99.7\%
CL intervals) of the PPDs of the parameters \lth, \lph, and
\ltp, for the \UpsOne, \UpsTwo, and \UpsThree\ states, in the HX frame.
Similar values are obtained in the 0.6--1.2 rapidity range~\cite{bib:SuppMaterial}.
Figure~\ref{fig:lambdatilde} displays the corresponding results for the frame-invariant
parameter \ltilde, including also the CS and PX values.
The results obtained
in the three frames are in good agreement, as required in the absence of
unaccounted for systematic effects.
Complete tables of results
for \lth, \lph, \ltp, and \ltilde, for the three \Ups\ states and in the three frames considered
in this analysis, are available in Ref.~\cite{bib:SuppMaterial}.

All the polarization parameters are compatible with zero or small values in the three
polarization frames, excluding that a significant polarization could remain undetected because of
smearing effects induced by unfortunate frame choices. The indication that the \UpsN\
resonances are produced as an unpolarized mixture might be related to the fact that the
measurements do not distinguish directly produced \Ups\ mesons from those produced in the
decays of heavier (P-wave) bottomonium states.

In summary, the polarizations of the \UpsN\
mesons produced in \Pp\Pp\ collisions at \mbox{$\sqrt{s} = 7$\TeV} have been determined as a
function of the \Ups\ \pt\ in two rapidity ranges and in three different polarization frames, using both
frame-dependent and frame-independent parameters.
The results exclude large transverse or longitudinal \UpsN\ polarizations, beyond the \pt\ and $y$
ranges probed by previous experiments, especially for the \UpsThree\ state, less
affected by feed-down decays, and are in disagreement with theoretical expectations for
high-energy hadron collisions~\cite{NLO-NRQCD,NNLO-CSM}.

We congratulate our colleagues in the CERN accelerator departments for the excellent performance of the LHC machine. We thank the technical and administrative staff at CERN and other CMS institutes, and acknowledge support from: BMWF and FWF (Austria); FNRS and FWO (Belgium); CNPq, CAPES, FAPERJ, and FAPESP (Brazil); MES (Bulgaria); CERN; CAS, MoST, and NSFC (China); COLCIENCIAS (Colombia); MSES (Croatia); RPF (Cyprus); MoER, SF0690030s09 and ERDF (Estonia); Academy of Finland, MEC, and HIP (Finland); CEA and CNRS/IN2P3 (France); BMBF, DFG, and HGF (Germany); GSRT (Greece); OTKA and NKTH (Hungary); DAE and DST (India); IPM (Iran); SFI (Ireland); INFN (Italy); NRF and WCU (Korea); LAS (Lithuania); CINVESTAV, CONACYT, SEP, and UASLP-FAI (Mexico); MSI (New Zealand); PAEC (Pakistan); MSHE and NSC (Poland); FCT (Portugal); JINR (Armenia, Belarus, Georgia, Ukraine, Uzbekistan); MON, RosAtom, RAS and RFBR (Russia); MSTD (Serbia); SEIDI and CPAN (Spain); Swiss Funding Agencies (Switzerland); NSC (Taipei); TUBITAK and TAEK (Turkey); STFC (United Kingdom); DOE and NSF (USA).

\bibliography{auto_generated}   % will be created by the tdr script.

\cleardoublepage \appendix\section{The CMS Collaboration \label{app:collab}}\begin{sloppypar}\hyphenpenalty=5000\widowpenalty=500\clubpenalty=5000\textbf{Yerevan Physics Institute,  Yerevan,  Armenia}\\*[0pt]
S.~Chatrchyan, V.~Khachatryan, A.M.~Sirunyan, A.~Tumasyan
\vskip\cmsinstskip
\textbf{Institut f\"{u}r Hochenergiephysik der OeAW,  Wien,  Austria}\\*[0pt]
W.~Adam, E.~Aguilo, T.~Bergauer, M.~Dragicevic, J.~Er\"{o}, C.~Fabjan\cmsAuthorMark{1}, M.~Friedl, R.~Fr\"{u}hwirth\cmsAuthorMark{1}, V.M.~Ghete, J.~Hammer, N.~H\"{o}rmann, J.~Hrubec, M.~Jeitler\cmsAuthorMark{1}, W.~Kiesenhofer, V.~Kn\"{u}nz, M.~Krammer\cmsAuthorMark{1}, I.~Kr\"{a}tschmer, D.~Liko, I.~Mikulec, M.~Pernicka$^{\textrm{\dag}}$, B.~Rahbaran, C.~Rohringer, H.~Rohringer, R.~Sch\"{o}fbeck, J.~Strauss, A.~Taurok, W.~Waltenberger, C.-E.~Wulz\cmsAuthorMark{1}
% ...
\vskip\cmsinstskip
\dag:~Deceased\\
1:~~Also at Vienna University of Technology, Vienna, Austria\\
% ...

\end{sloppypar}
\end{document}